\documentclass[12pt]{iopart}

\usepackage{iopams}
\usepackage[dvips]{graphicx}
\begin{document}

\title{Analytic Solution to Clustering Coefficients on Weighted Networks}

\author{Yichao Zhang}\address{Department of Computer Science and Technology, Tongji
University, 4800 Cao'an Road, 201804, Shanghai, China}
\author{Zhongzhi Zhang}\address{School of Computer Science, Fudan University, 200433,
Shanghai, China\\Shanghai Key Lab of Intelligent Information
Processing, Fudan University, 200433, Shanghai, China}
\author{Jihong Guan}\ead{jhguan@mail.tongji.edu.cn}\address{Department of Computer Science and Technology, Tongji University,
4800 Cao'an Road, 201804, Shanghai, China}
\author{Shuigeng Zhou}\address{School of Computer Science, Fudan University, 200433,
Shanghai, China\\Shanghai Key Lab of Intelligent Information
Processing, Fudan University, 200433, Shanghai, China}

\begin{abstract}
Clustering coefficient is an important topological feature of
complex networks. It is, however, an open question to give out its
analytic expression on weighted networks yet. Here we applied an
extended mean-field approach to investigate clustering coefficients
in the typical weighted networks proposed by Barrat, Barth\'{e}lemy
and Vespignani (BBV networks). We provide analytical solutions of
this model and find that the local clustering in BBV networks
depends on the node degree and strength. Our analysis is well in
agreement with results of numerical simulations.
\end{abstract}

\maketitle

\section{Introduction}
The concept of weight plays an important role in characterizing the
features of complex systems in nature, since many different
correlations in the real world can be represented by weighted links.
In the past decade, as an emerging network dynamics, weight dynamics
has attracted a large number of scientific
communities~\cite{PNAS1013747,ARIST41537,PRL94188702,PRE71066124,PRE72017103,PRE74036111,PA378591,PRE75026111,PRE72046140,PRE73016133}.
It has been shown that many biological, technological, and social
systems are best described by weighted networks, as the nodes
interact with each other with varying strength. To describe the
variety, a link between two nodes is associated with a weight in the
weighted network. The weight of a link can represent different
properties for different networks. For a social network, link weight
can indicates the intimacy between two
individuals~\cite{PNAS98404,PRE64016131}; for the Internet, link
weight can represent the bandwidth of an optical cable between two
routers~\cite{NATURE401130}; for the world-wide aviation network,
link weights can reflect the annual volume of passengers traveling
between two airports~\cite{PNAS1013747}; and for the \emph{E.~coli}
metabolic network, link weight can encode the optimal metabolic
fluxes between two metabolites~\cite{Nature427839}.

Previous studies have pointed out the importance of the link weight
of networks and led to the formulation of a long list of models
aimed at incorporating the characteristics in weighted
networks~\cite{PNAS1013747,ARIST41537,PRL94188702,PRE71066124,PRE72017103,PRE74036111,PA378591,PRE75026111,PRE72046140,PRE73016133,PRE71026103,PRE72056138,PRE73025103,PRE73056109,PRE70066149}.
Among these models, the one proposed by Barrat, Barth\'{e}lemy and
Vespignani (BBV)~\cite{PNAS1013747,PRE70066149,PRL92228701} has been
deemed to be the pioneer and classic work by most physics
communities. The main contribution of their work can be ascribed to
two aspects: a new strength driven preferential attachment
mechanism; local weight and strength rearrangements for a new link.
The results of this model represent many behaviors observed
empirically in real-world networks, e.g., link weight-degree
correlation, node strength-degree correlation, and scale-free
topology.

Currently, a primary purpose of the study of complex networks is to
understand how their dynamical behaviors are influenced by
underlying geometrical and topological properties~\cite{PR424175,
SIAMR45167}. Among many fundamental structural
characteristics~\cite{AP56167}, clustering coefficient, also known
as transitivity, is an important topological feature of complex
networks, generating considerable attention. In the sociological
literature, for example, it is referred to as the ``network
density". For node $i$ with $k_i$ links, $C_i$ denote the fraction
of these allowable edges that actually exist. The local clustering
coefficient for the whole system is given as the average of $C_i$
for each vertex~\cite{NATURE393440}. As is known, average local
clustering coefficient is one of the two key criterions to judge
whether a network or a complex system possess the small-world
property. Most previous related studies have been confined to
numerical methods, which is, however, prohibitively difficult for
large networks because of the limit of time and memory. Hence, there
is a need to give out a analytical solution of the clustering
coefficient toward providing a useful insights into topological
structures of complex networks.

Despite the necessity, to the best of our knowledge, the rigorous
computation for the clustering coefficient of BBV model has not been
addressed, contrasting sharply with the successes in other network
properties such as degree, weight and strength
distribution~\cite{PRE70066149}. To fill this gap, in this present
paper we investigate this interesting quantity analytically. We
derive an exact formula for the clustering coefficient
characterizing the BBV model. The analytic method we adopt in the
work, is an extended mean field approach. The obtained precise
result shows that the average local clustering coefficient of BBV
network has a power-law decrease with the number of nodes. To some
extent, our research opens a way to theoretically study the
clustering coefficient of neutral or nonassortative
mixing~\cite{PRL89208701} weighted networks on the one hand. On the
other hand, our exact solution may give insight different from that
afforded by the approximate solution of numerical simulation.

\section{\label{sec:BBV} Introduction to the model}
The BBV network starts from an initial seed of $N_0$ vertices fully
connected by links with assigned weight $w_0$. A new vertex $n$ is
added at each time step. This new site is connected to $m$
previously existing vertices (i.e., each new vertex will have
initially exactly $m$ edges, all with equal weight $w_0$), choosing
preferentially sites with large strength; i.e., a node $i$ is chosen
according to the probability
\begin{equation}
\Pi_{n\rightarrow i}=\frac{s_i}{\Sigma_v s_v}.\label{pi}
\end{equation}
where $s_i$ is the strength of the node $i$, defined as $s_i=\sum_j
w_{ij}$. The weight of each new link $L_{ij}$, is initially set to a
given value $w_0$. The creation of this edge will introduce a local
variations of the edge weight $w_{ij}\rightarrow
w_{ij}+\Delta_{w_{ij}}$, in which node $j$ is one of node $i$'s
neighbors and node strength of node $s_j\rightarrow
s_j+\Delta_{w_{ij}}$, where this perturbation is proportionally
distributed among the edges according to their weights.
\begin{equation}
\Delta_{w_{ij}}=\delta_i \frac{w_{ij}}{s_i}.
\end{equation}

Simultaneously, this rule yields a total strength increase for node
$i$ of $\delta_i +w_0$, implying that $s_i\rightarrow s_i+ \delta_i
+w_0$. In this paper, we focus on the case of homogeneous coupling
with $\delta_i=\delta=const$ and equivalently set $w_0=1$. After the
weights and strengths have been updated, the growth process is
iterated by introducing a new vertex, until the desired size of the
network is reached.

In BBV networks, the dynamical process for $s_i$ and $k_i$ can thus
be described by the following evolution:
\begin{equation}\label{evolution_s_i}
\frac{ds_i}{dt}=m\frac{s_i(t)}{\Sigma_v
s_v}(1+\delta)+\Sigma_{j\in\Omega(i)}m\frac{s_i(t)}{\Sigma_v
s_v}\delta\frac{w_{ij}(t)}{s_j(t)},
\end{equation}
\begin{equation}\label{evolution_k_i}
\frac{dk_i}{dt}=m\frac{s_i(t)}{\Sigma_v s_v}.
\end{equation}
Note that $\sum_{j\in\Omega(i)}$ runs over the nearest neighbors of
the node $i$. Solving these two equations with initial conditions
$k_i(t_i)=s_i(t_i)=m$, one can have
\begin{equation}
s_i=m\left(\frac{t}{t_i}\right)^{\frac{2\delta+1}{2\delta+2}}\label{si}
\end{equation}
~\cite{PRE70066149}, and
\begin{equation}
k_i=\frac{s_i+2m\delta}{2\delta+1}\label{ki}.
\end{equation}

After introducing the BBV model, we now investigate analytically the
clustering coefficient of all the nodes in the BBV networks. As is
known, like BA networks, the clustering coefficient in BBV networks
rapidly decreases with the network size $t$. In the previous
studies, researchers give out numerical calculations to show this
mechanism while we will offer an exact solution to explain it in
what follows.
\begin{figure}
\begin{center}
\scalebox{0.3}[0.3]{\includegraphics{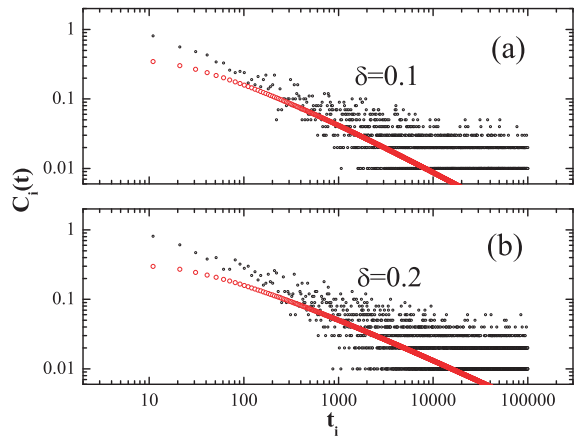}}\caption{(Color
online) The initial value of the local clustering coefficient
$C_{i}(t_{i})$ for (a) $\delta=0.1$ and (b) $\delta=0.2$, (averaged
over $10$ BBV networks). Red circles show the analytical solutions
given by Eq.~(\ref{citi}). Note that the $C_{i}(t)$ axis is
nonlinear.}\label{fig_citi}
\end{center}
\end{figure}

\section{\label{sec:calculation} Analytical expression to clustering coefficients of the model}
Before introducing the way to calculate the clustering coefficient,
it is necessary to build up a conception, `what is neutral weighted
networks', at first.

In networks, the degree of a node is defined as the number of links
the node has. A relevant topological property is the degree-degree
correlation, or so-called assortative mixing. It is defined as a
preference for high-degree vertices to attach to other high-degree
vertices and vice versa~\cite{PRL89208701,pre67026126}. By contrast,
disassortative mixing is defined as a preference for high degree
vertices preferentially connect with low degree ones and vice versa.
However, a network with neutral degree correlation means neither
assortative nor disassortative. A random network, where links are
randomly connected between nodes, is an example of neutral networks.

Along with the degree correlations, we introduce the weighted degree
correlations~\cite{PNAS1013747}, defined as
\begin{equation}
k^w_{\rm nn,i}=\frac{1}{s_i}\sum^{N}_{j=1}a_{ij}w_{ij}k_j.
\end{equation}
The $k^w_{\rm nn,i}$ thus measures the effective affinity to connect
with high- or low-degree neighbors according to the magnitude of the
actual interactions. As well, the behavior of the function
$k^w_{nn}(k)$ marks the weighted assortative or disassortative
properties. When $k^w_{\rm nn}(k)$ increases with $k$, it means that
nodes have a tendency to connect to nodes with a similar or larger
degree. In this case the network is defined as assortative. In
contrast, if $k^w_{\rm nn}(k)$ is decreasing with $k$, which implies
that nodes of large degree are likely to have near neighbors with
small degree, then the network is said to be disassortative.
Naturally, if correlations are absent, it can be ascribed to
weighted neutral networks.

Of particular interest is the BBV model mentioned in
section~\ref{sec:BBV} can give rise to neutral networks, for small
$\delta$~\cite{PRE70066149}. However, for increasing $\delta$, the
disassortative character and a power-law behavior of $k_{\rm nn}(k)$
emerge. Fortunately, their weighted assortativity $k^w_{\rm nn}(k)$
is a constant, that is, they belong to weighted neutral networks. In
this case, we redefine the linking probability of node $i$ and $j$,
\begin{equation}\label{pij}
p_{ij}=\frac{s_js_i}{\sum_{v}s_{v}}=\frac{s_js_i}{2m(1+\delta)tW_{ij}},
\end{equation}
for the sake of extending the mean field method to this particular
network for large $\delta$.
\begin{figure}
\begin{center}
\scalebox{0.8}[0.8]{\includegraphics{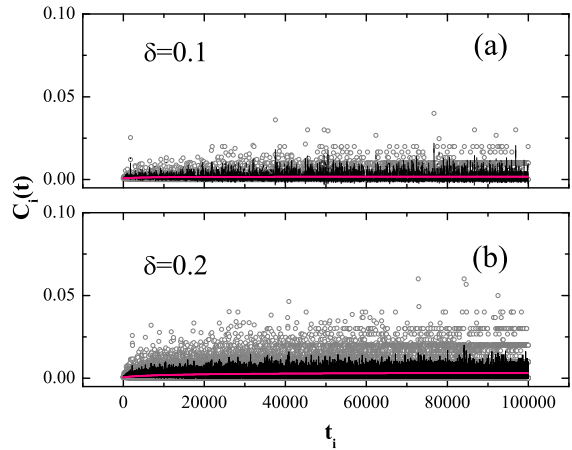}}\caption{(Color
online) The local clustering coefficient $C_{i}(t)$ as a function
$t_{i}$ for (a) $\delta=0.1$ and (b) $\delta=0.2$. Results are
obtained by numerical simulations on BBV networks (averaged over 10
networks). Peachblow lines show the analytical solutions given by
equation~(\ref{cit}), and black curves are the smoothed data of
simulation data (gray circles).} \label{fig_cit}
\end{center}
\end{figure}

In a BBV network, consider consequently a certain node $i$ at time
step $t$. Note that we only consider the case $m>1$ in that the BBV
network is close to tree-like networks when $m=1$. As is known to us
all, for a tree network, the clustering coefficient in the network
is equal to zero. By the definition~\cite{NATURE393440}, the
clustering coefficient $C_{i}$ depends on two variables $E_{i}$ and
$k_{i}$. Its definition is $C_i=\frac{2E_i}{k_i(k_i-1)},$ where
$E_i$ is the number of links between the $k_i$ neighbors of $i$.

Since in the BBV model only new nodes may create links, the
coefficient $C_{i}$ changes only when its degree $k_{i}$ changes,
i.e., when new nodes create connections to $i$ and $x\in\langle
0,m-1\rangle$ of its nearest neighbors. The appropriate equation for
changes of $C_{i}$ is then
\begin{equation}\label{dcidt}
\frac{dC_{i}}{dt}=\sum_{x=0}^{m-1}\widetilde{p}_{ix}\Delta_{C_{ix}},
\end{equation}
where $\Delta_{C_{ix}}$ denotes the change of the clustering
coefficient when a new node connects to the node $i$ and to $x$ of
its first neighbors as well, whereas $\widetilde{p}_{ix}$ describes
the probability of this event.

Notice that, in the equation~(\ref{dcidt}), $\Delta_{C_{ix}}$
represents the difference between clustering coefficients of the
same node $i$ calculated after and before a new node attachment
\begin{equation}\label{eq1a}
\Delta_{C_{ix}}=\frac{2(E_{i}+x)}{k_{i}(k_{i}+1)}-\frac{2E_{i}}{k_{i}(k_{i}-1)}=
\frac{2x}{k_{i}(k_{i}+1)}-\frac{2C_{i}}{k_{i}+1},
\end{equation}
where $\Delta_{C_{ix}}>0$. Also, the equation~(\ref{eq1a}) gives out
the restricted region of our method. The probability
$\widetilde{p}_{ix}$ is a product of two factors.
\begin{equation}\label{pix}
\widetilde{p}_{ix}=\frac{s_i}{\Sigma_v
s_v}\left(^{m-1}_{\:\:\:\:x}\right) P^{x}(1-P)^{m-1-x}.
\end{equation}
The first factor is the probability of a new link to end up in $i$,
which is given by equation~(\ref{pi}). The second one is the
probability that $x$ among the rest of $(m-1)$ new links connect to
the nearest neighbors of $i$. It is equivalent to the probability
that $(m-1)$ Bernoulli trials with the probability $P$ for $x$
successes. So that,
\begin{equation}\label{P}
P=\frac{\sum_{j\in\Omega(i)}s_{j}}{\sum_{v}s_{v}}=\frac{\sum_{j\in\Omega(i)}s_{j}}{2m(1+\delta)t}.
\end{equation}
Replacing the sum $\sum_{j\in\Omega(i)}$ by an integral, one obtains
\begin{equation}\label{sj+}
\sum_{j\in\Omega(i)}s_{j}\sim \int_{1}^{t}s_{j}\:p_{ij}\:dt_{j}.
\end{equation}
where,
\begin{equation}
p_{ij}=\frac{s_is_j}{\sum_{v}s_{v}W_{ij}}=\frac{s_is_j}{2m(1+\delta)tW_{ij}},
\end{equation}
in which $W_{ij}$ is the weight of the link $L_{ij}$. In BBV
networks,
\begin{equation}
W_{ij}(t)=\left(\frac{t}{t_{ij}}\right)^{\frac{\delta}{\delta+1}},
\end{equation}
where $t_{ij}=\max(i,j)$. Hence, using the equation~(\ref{sj+}) and
one can find the probability
\begin{equation}\label{sj++}
P=
\frac{mt^{\frac{4\delta-1}{2\delta+2}}}{4(1+\delta)\delta}\left(i^{-\frac{4\delta+1}{2\delta+2}}-\frac{i^{-\frac{6\delta+1}{2\delta+2}}}{2}-\frac{t^{-\frac{2\delta+1}{2\delta+2}}i^{-\frac{2\delta+1}{2\delta+2}}}{2}\right).
\end{equation}
Now, inserting equation~(\ref{si}), (\ref{ki}), (\ref{eq1a}) and
(\ref{pix}) into equation~(\ref{dcidt}) one can obtain
\begin{eqnarray}\label{eq2}
&\frac{dC_{i}}{dt}+\frac{m(2\delta+1)t^{\frac{-1}{2\delta+2}}C_{i}}{(1+\delta)[mt^{\frac{2\delta+1}{2\delta+2}}+(2m\delta+2\delta+1)t_i^{\frac{2\delta+1}{2\delta+2}}]}&\nonumber\\
&=\frac{t^{\frac{-3}{2\delta+2}}}{2\delta(\delta+1)^2}\cdot\frac{m(m-1)(2\delta+1)}{mt^{\frac{2\delta+1}{2\delta+2}}+(2m\delta+2\delta+1)t^{\frac{2\delta+1}{2\delta+2}}_i}.&
\end{eqnarray}
In the derivation process, we simplify
\begin{equation}\label{prox}
\frac{mt^{\frac{2\delta+1}{2\delta+2}}+(2m\delta+2\delta+1)i^{\frac{2\delta+1}{2\delta+2}}}{mt^{\frac{2\delta+1}{2\delta+2}}+2m\delta
i^{\frac{2\delta+1}{2\delta+2}}}\approx1
\end{equation} to make the analytical integral easily executed. Solving the
equation for $C_{i}$ one gets
\begin{eqnarray}\label{cit}
&C_i(t)=\left[mt^{\frac{2\delta+1}{2\delta+2}}+(2m\delta+2\delta+1)t^{\frac{2\delta+1}{2\delta+2}}_i\right]^{-2}\cdot&\nonumber\\
&\left[B_i+\frac{m^2}{4(1+\delta)\delta^2}\left(\frac{2t^{\frac{3\delta}{\delta+1}}t^{-\frac{\delta}{\delta+1}}_i-t^{\frac{3\delta}{\delta+1}}t^{-\frac{2\delta}{\delta+1}}_i-6t^{\frac{\delta}{\delta+1}}}{6}\right)
\right],\ \ &
\end{eqnarray}
where $B_i$ is an integration constant for the node $i$ and
determined by the initial condition $C_{i}(t_{i})$ that describes
the clustering coefficient of the node $i$ exactly at the moment of
its attachment $t_{i}$. Note that $\Delta_{C_{ix}}>0$ in
equation~(\ref{eq1a}), we have $\frac{1}{k_i}>C_i$. Thus, in this
article, it is worth noticing that $\delta$ has to be confined in
the region $[0,1/4)$. The restriction is given by our approximation
in equation~(\ref{prox}). We believe this drawback will be addressed
in future work.

To clarify the calculation of $C_i(t_{i})$, we show another form of
the definition of $C_i$~\cite{SIAMR45167,NATURE393440},
$$C_i=\frac{number\ of\ triangles\ connected\ to\ vertex\ i}{number\
of\ triples\ centered\ on\ vertex\ i}.$$ It can be rewritten as
\begin{eqnarray}\label{citi}
C_{i}(t_{i})=\frac{\sum_{j}\sum_{v}p_{ij}\:p_{iv}\:p_{jv}\:}{2\left(^{m}_{\:2}\right)}
\sim\frac{m^{2}\left(1-t_i^{\frac{-\delta}{\delta+1}}\right)\left(1-t_i^{\frac{-2\delta}{\delta+1}}\right)t_i^{\frac{\delta-1}{\delta+1}}}{16(m-1)(\delta+1)\delta^2}.
\end{eqnarray}

Figure~\ref{fig_citi} shows the prediction of the
equation~(\ref{citi}) in comparison with numerical results. For
small values of $t_{i}$ the numerical data differ from the theory
significantly. This incongruity is caused by the formula for the
probability of a connection $p_{ij}$ in equation~(\ref{pij}) used
three times in equation~(\ref{citi}), which holds only in the
asymptotic region $t_{i}\longrightarrow\infty$.

\begin{figure}
\begin{center}
\scalebox{0.8}[0.8]{\includegraphics{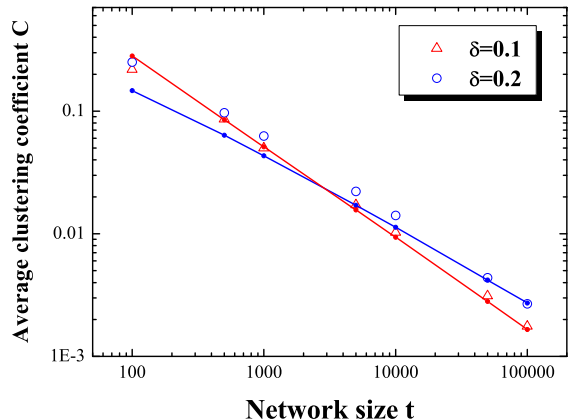}}\caption{(Color online)
The clustering coefficient $C$ of a whole BBV network as a function
of the network size $t$ for different values of the $\delta$.
Results are obtained by numerical simulations on BBV networks
(averaged over 10 networks). The solid lines are numerical solutions
given by equation~(\ref{numC}).}\label{fig_avc}
\end{center}
\end{figure}

Taking into account the initial condition $C_i(t_i)$ from
equation~(\ref{citi}), one can obtain $B_i$ of a given node $i$
\begin{eqnarray}\label{cit1}
&B_i=\frac{m^{2}(2m\delta+2\delta+m+1)^2\left(1-t_i^{\frac{-\delta}{\delta+1}}\right)\left(1-t_i^{\frac{-2\delta}{\delta+1}}\right)t_i^{\frac{3\delta}{\delta+1}}}{16(m-1)(\delta+1)\delta^2}&\nonumber\\
&-\frac{m^2}{4(1+\delta)\delta^2}\left(\frac{t_i^{\frac{2\delta}{\delta+1}}}{3}-\frac{7t_i^{\frac{\delta}{\delta+1}}}{6}\right).&
\end{eqnarray}
To obtain the average clustering coefficient $C$ of the whole
network, the expression equation~(\ref{cit1}) has to be averaged
over all nodes within a network. Admittedly, it is difficult to find
an exact analytic result of $C$ uniformly but we can give out
numerical solutions of BBV networks by the definition
\begin{equation}\label{numC}
C=\frac{\sum_{1}^{t}C_{i}(t)}{t}.
\end{equation}
In the figure~(\ref{fig_avc}), one can easily find out that both the
numerical and theoretical results exhibit a power-law decay as the
size of networks growing. Hence, when $t$ tends to infinity, $C$ is
getting close to zero. As shown in the figure, the power-law
exponent depends on the the tunable parameter $\delta$. For large
$\delta$, the decreasing velocity is relatively slow.

\clearpage

\section{\label{sec:conclusion}Conclusion\protect}
To sum up, the clustering coefficient in complex systems is an
important topological feature of complex systems. It has a profound
impact on a variety of crucial fields, such as epidemic processes,
network resilience and finding community structure, to name but a
few. In this paper, we have derived analytically the solution for
the clustering coefficient of the Barrat et al.'s
model~\cite{PRE70066149} which has been attracting much research
interest. We found that the average clustering coefficient decays
with the size of networks growing, following a power-law. Our
analytical technique could guide and shed light on related studies
for weighted networks by providing a paradigm for calculating the
clustering coefficient.


\section*{References}
\begin{thebibliography}{27}
\bibitem{PNAS1013747} A. Barrat, M. Barth\'{e}lemy, R. Pastor-Satorras, and A. Vespignani, Proc. Natl. Acad. Sci. U.S.A. {\bf 101}, 3747 (2004).
\bibitem{ARIST41537} K. Borner, S. Sanyal, and A. Vespignani, Ann. Rev. Infor. Sci. Tech. {\bf 41}, 537 (2007).
\bibitem{PRL94188702} W. X. Wang, B. H. Wang, B. Hu, G. Yan, and Q. Ou, Phys. Rev. Lett. {\bf 94}, 188702 (2005).
\bibitem{PRE71066124} Z. X. Wu, X. J. Xu, and Y. H. Wang, Phys. Rev. E {\bf 71}, 066124 (2005).
\bibitem{PRE72017103} K. I. Goh, B. Kahng, and D. Kim, Phys. Rev. E {\bf 72}, 017103 (2005).
\bibitem{PRE74036111} G. Mukherjee and S. S. Manna, Phys. Rev. E {\bf 74}, 036111 (2006).
\bibitem{PA378591} C. C. Leung, H. F. Chau, Physica A {\bf 378}, 591 (2007).
\bibitem{PRE75026111} Y. B. Xie, W. X. Wang, and B. H. Wang, Phys. Rev. E {\bf 75}, 026111 (2007).
\bibitem{PRE72046140} W. X. Wang, B. Hu, T. Zhou, B. H. Wang, and Y. B. Xie, Phys. Rev. E {\bf 72}, 046140 (2005).
\bibitem{PRE73016133} W. X. Wang, B. Hu, B. H. Wang, and G. Yan, Phys. Rev. E {\bf 73}, 016133 (2005).
\bibitem{PNAS98404} M. E. J. Newman, Proc. Natl. Acad. Sci. U.S.A. {\bf 98}, 404 (2001).
\bibitem{PRE64016131} M. E. J. Newman, Phys. Rev. E {\bf 64}, 016131 (2001).
\bibitem{NATURE401130} R. Albert, H. Jeong, and A.-L. Barab\'{a}si, Nature {\bf 401}, 130 (1999).
\bibitem{Nature427839} E. Almaas, B. Kov\'{a}cs, Z. N. Oltval, and A.-L. Barab\'{a}si, Nature {\bf 427}, 839 (2004).
\bibitem{PRE71026103} T. Antal and P. L. Krapivsky, Phys. Rev. E {\bf 71}, 026103 (2005).
\bibitem{PRE72056138} P. V. Mieghem and S. M. Magdalena, Phys. Rev. E {\bf 72}, 056138 (2005).
\bibitem{PRE73025103} T. Kalisky, S. Sreenivasan, L. A. Braunstein, S. V. Buldyrev, S. Havlin, and H. E. Stanley, Phys. Rev. E {\bf 73}, 025103 (2006).
\bibitem{PRE73056109} Z. Pan, X. Li, and X. Wang, Phys. Rev. E {\bf 73}, 056109 (2006).
\bibitem{PRE70066149} A. Barrat, M. Barth\'{e}lemy, and A. Vespignani, Phys. Rev. E {\bf 70}, 066149 (2004).
\bibitem{PRL92228701} A. Barrat, M. Barth\'{e}lemy, and A. Vespignani, Phys. Rev. Lett. {\bf 92}, 228701 (2004).
\bibitem{PR424175} S. Boccaletti, V. Latora, Y. Moreno, M. Chavezf, and D. U. Hwanga, Physics Report {\bf 424}, 175 (2006).
\bibitem{SIAMR45167} M. E. J. Newman, SIAM Rev. {\bf 45}, 167 (2003).
\bibitem{AP56167} L. da. F. Costa, F. A. Rodrigues, G. Travieso, and P. R. V. Boas, Adv. Phys. {\bf 56}, 167 (2007).
\bibitem{NATURE393440} D. J. Watts and S. H. Strogatz,  Nature (London) {\bf 393}, 440 (1998).
\bibitem{PRL89208701} M. E. J. Newman, Phys. Rev. Lett. {\bf 89}, 20 (2002).
\bibitem{pre67026126} M. E. J. Newman, Phys. Rev. E {\bf 67}, 026126 (2003).
\end{thebibliography}
\end{document}